# Optical Magnetometer: Quantum Resonances at pumping repetition rate of 1/n of the Larmor frequency

Andrei Ben Amar Baranga, Alexander Gusarov, Gennady A. Koganov, David Levron and Reuben Shuker*

Physics Department, Ben Gurion University of the Negev, Beer Sheva 8410501, Israel

*shuker@bgu.ac.il

**Abstract**

The response of a SERF atomic magnetometer to a repetitive short-pulsed pump was investigated. Quantum sub-resonances at a repetition rate of 1/n of the Larmor frequency of the magnetic field inside the shield are experimentally observed and theoretically explained. This is a type of synchronization phenomenon. Investigations in single alkali atoms cells as well as mixed alkali atoms of K and Rb are presented. In the later, one species is pumped while the probe is on the other specie polarized by spin exchange. The effect of spin destruction, spin exchange and collisions are studied in order to account for the width of the resonances. Quantum calculations of a three levels $\Lambda$ model for this phenomenon exhibit a dip at the resonance frequency in the absorption spectrum for both cases of pulsed and CW pump modes and an evidence for EIT.

**Introduction**

Atomic magnetometers are studied vigorously in the last decades since a record sensitivity below 1 fT was achieved in a SERF configuration [1]. Various configurations of optical magnetometers as well as their applications have been published [2,3] along and in comparison with non-optical highly sensitive magnetometers [4]. A dual use, scalar and vectorial optical magnetometer [5] and an unshielded pulsed magnetometer [6] increase their applications' span.

In one configuration of an atomic magnetometer, alkali atoms in a glass cell are optically pumped by a circularly polarized laser beam. A magnetic field rotates the polarized atoms at Larmor frequency to be depicted by a linearly polarized probe beam. Commonly, the pump and the probe lasers are CW. In a SERF configuration of an atomic magnetometer the cell is warmed up to increase alkali density and it is magnetically shielded to reduce to minimum any external magnetic fields to achieve its record sensitivity.

In this article, we present a particular effect of quantum resonances achieved in a SERF atomic magnetometer optically pumped by repetitive short-pulses, in which the repetition rate is 1/n the atom internal Larmor frequency. Experimental and theoretical investigations are presented.

## Experimental

In this experiment, we used a pulsed mode of optical pumping. Pulses with duration of less than 1 msec at a variable repetition rate are used. The schematic of the pulsed operation of a SERF magnetometer is shown in Fig. 1. We used a hybrid magnetometer configuration with a glass cell containing a mixture of Rb and K alkali atoms, 52 Torr of $N_2$ and 2.5 amagats He buffer gas. The cell and oven are placed inside a five-layer µ-metal magnetic shields. The cell is heated to 180°C by hot air flow. At this temperature the ratio of Rb/K densities is 1/20. Magnetic fields inside the shields were nulled by a system of 3 pairs of compensation coils, producing calibrated, uniform magnetic fields along **x**, **y**, and **z** directions.

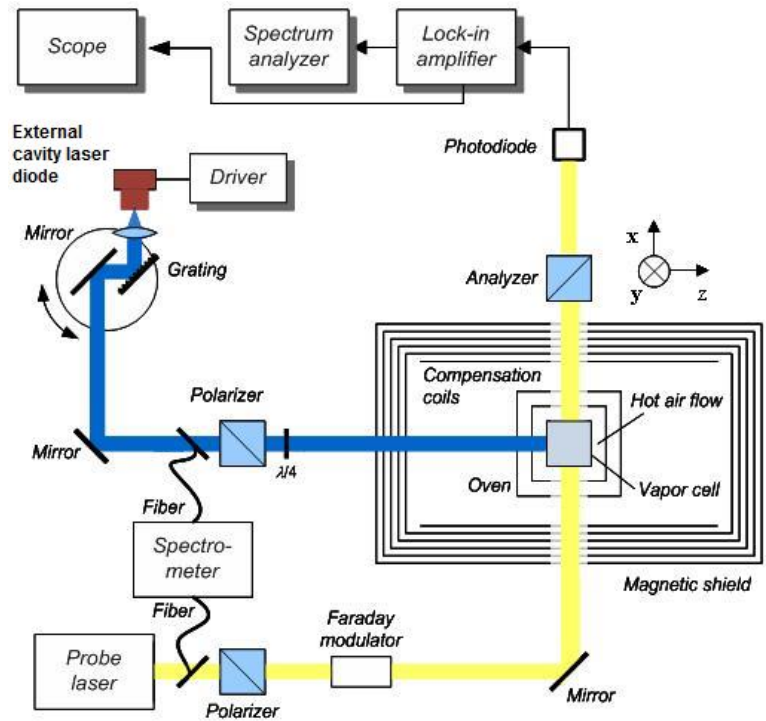

Fig. 1. Experimental set-up.

In a hybrid configuration one separates the pump from the probe/detection processes by pumping one species only. By optical pumping, an extended Zeeman level state is populated in a process involving many cycles of excitations, collisions, decays and thermal noise. This is followed by spin exchange to atoms of the other species, that encounters only direct exchange of the spin, avoiding many noise related effects. Probing is done on the second alkali. Thus, reduction of the "noise" on the detection signal is obtained, as evidenced by our experimental results and mentioned elsewhere [7,8].

Optical pumping is accomplished by circularly polarized laser light propagating in the **z**-direction tuned to the center of the D1 line ($n^2S_{1/2}$-$n^1P_{1/2}$) of the pumped alkali atoms by an external cavity Littrow configuration. The pump has been performed by an "*Intense*" CW Laser Diode. However, the diode is adapted to pulse mode and driven by DEI PCO-6141 high power pulsed current source to pulses of 200µs and 70µs durations. The linearly polarized probe beam generated by New Focus Velocity tunable diode laser, propagates in the **x**-direction. It is detuned by ~0.8nm from the D1 or D2 line ($m^2S_{1/2}$-$m^1P_{3/2}$) of the probed alkali atom, the pumped one or the other species, to reduce absorption and depolarization. The polarization plane of the probe beam is modulated by a Faraday modulator at a frequency $\omega_m \approx 3.1$ kHz. The probe beam passes through the cell and an analyzer, and is detected by a ThorLabs DET36A photodiode. The photo diode is connected to a SRS SR850 digital Lock-in amplifier. The filtered signals are connected to an SRS SR760 spectrum analyzer and a LeCroy Wave-Runner 44XI 400MHz Oscilloscope for analysis.

The cell containing Rb and K metals allows several pump-probe configurations. As mentioned before, one species is pumped while the probe was on the same species or on the other one that is polarized by spin exchange. We investigate several options looking for better signal to noise measurements and narrower resonances. The configurations are:

1. Rb D1 atoms pumping and K D1 atoms probing,
2. K D1 atoms pumping and Rb D1 atoms probing,
3. Rb D1 atoms pumping and Rb D2 atoms probing,
4. K D1 atoms pumping and K D2 atoms probing.

For each configuration, the magnetometer lasers are re-tuned to the required wavelengths. A constant magnetic field corresponding to 100Hz to 220Hz Larmor frequencies was applied to the vapor cell. Fig. 2 presents a typical magnetometer response to a pump laser pulse displaying a decaying "sine" function oscillating at the Larmor frequency.

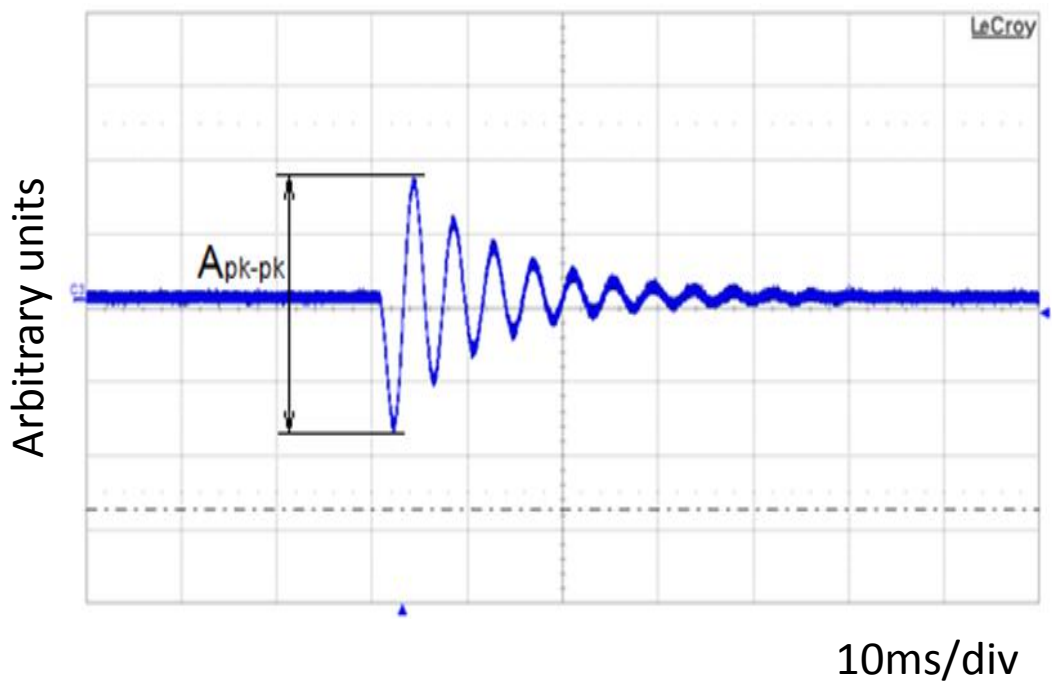

Fig. 2. Le Croy oscilloscope trace of a typical magnetometer response to a pumping laser pulse with a magnetic field applied inside the shield in *y* direction. The decaying sine function oscillates at the Larmor frequency of 220Hz.

Amplitude of the first oscillation A$_{pk-pk}$ (Fig. 2) of the magnetometer signal was measured as a function of the pumping repetition rate. The obtained resonance curves for the four different magnetometer configurations are summarized in Fig. 3, with a detailed description in figure caption. We found a synchronization between the Larmor frequency $\omega_L$ and the pump pulse repetition rate, *f*, leading to resonance features at $2\pi f = \omega_L/n$, where n is integer (inverse harmonics). The Larmor frequency is 220 Hz. Resonances at 220Hz, 110Hz, 73.3Hz, 55Hz, (1, 1/2, 1/3, 1/4… of the Larmor) are clearly observed corresponding to $f = \omega_L/n$, with n=1,2...

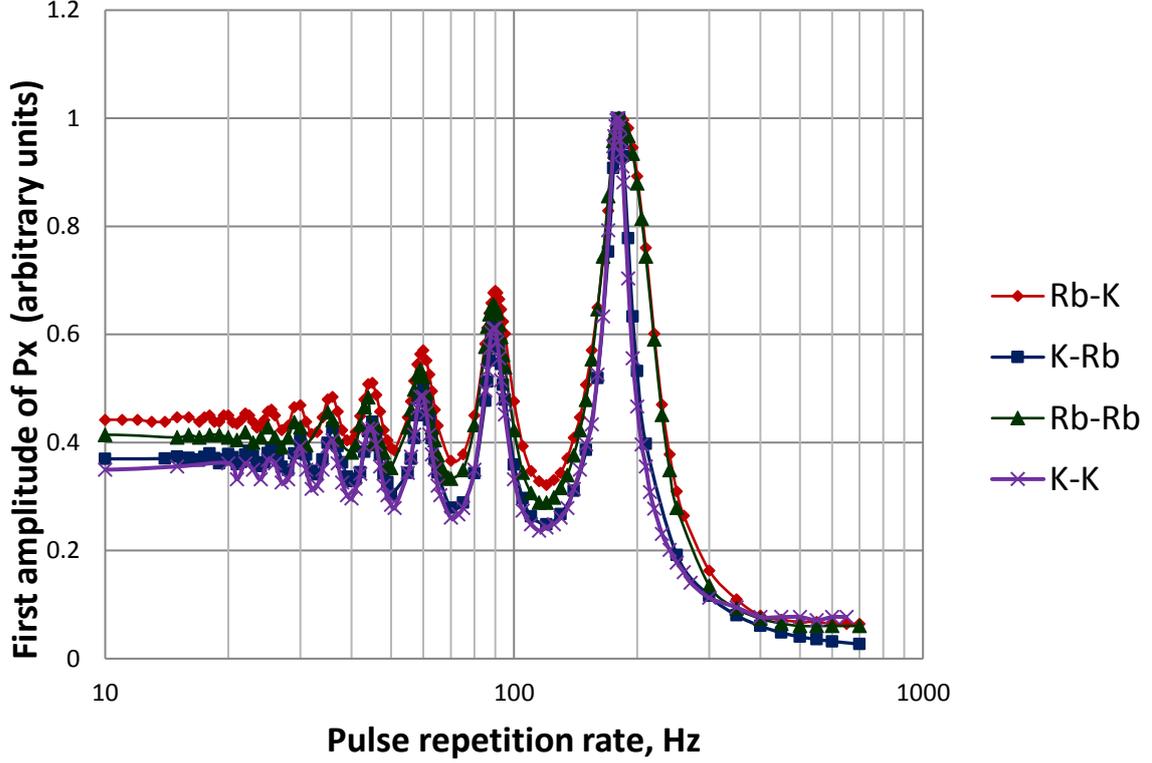

Fig. 3. Normalized resonance curves for 180Hz Larmor frequency for the different pump-probe magnetometer configurations. Magnetometer signal for Rb pumping is stronger, but the FWHM is wider for Rb pumped atoms at all the measured magnetic fields and narrower for potassium pumping.

Signals from an optically pumped atomic magnetometer are proportional to the x component of the spin polarization. We used Bloch equations describing the behavior of the spin polarization to simulate the behavior of these magnetometers. The Bloch equations for a hybrid magnetometer[8] using a cell containing high pressure (2.5 amagats) of He as buffer gas, K vapor and Rb vapor, after the pump pulse (OFF time) are given by Equations 1 and 2:

$$\frac{d}{dt}S^K = \gamma^K S^K \times B + R_{SE} S^{Rb} - (R_{SD}^K + R_{SE}^K)S^K \qquad (1)$$

$$\frac{d}{dt}S^{Rb} = \gamma^{Rb} S^{Rb} \times B + R_{SE} S^K - (R_{SD}^{Rb} + R_{SE}^{Rb})S^{Rb} \qquad (2)$$

Where $\gamma$ is the gyromagnetic ratio, $R_{SD}$ is the spin relaxation rate due to spin destruction collisions, and $R_{SE}$ is the spin relaxation rate due to spin-exchange collisions. The diffusion term is neglected because of the high pressure of the buffer gas. There exists an analytical solution to Eq. 1 and Eq. 2. Using the simple geometry depicted in Fig. 1, and in the simplest case, assuming initial conditions such that at the end of the pump pulse, the polarization of the 2$^{nd}$ species is zero and the polarization of the 1$^{st}$ species is in the x-axis direction, and further assuming that the spin exchange rate and the spin destruction rate are the same for the two species, the polarizations of the two species are given by Eq. 3 and Eq. 4 and depicted in Fig.4.

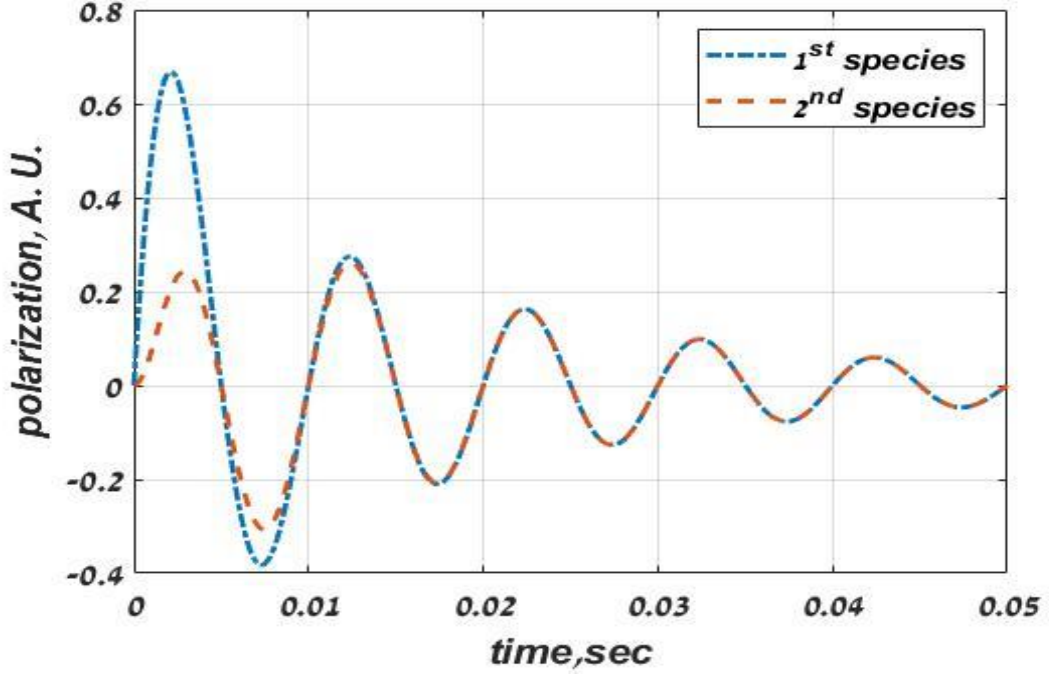

Fig. 4. Polarization transfer from species 1 to species 2 in a hybrid atomic magnetometer

$$P_{1x}=e^{-\Gamma t} \cosh(R_{SE+SD}t) \sin(\gamma B_z t) \qquad (3)$$

$$P_{2x}=e^{-\Gamma t} \sinh(R_{SE+SD}t) \sin(\gamma B_z t) \qquad (4)$$

A theoretical reconstruction of the resonance curve was obtained by solving the Bloch equation for a single short pulse of pump laser light, and recording the amplitude of the first oscillation after the end of the pump pulse. The single pump pulse was applied repetitively using the polarization values calculated from the n$^{th}$ application as initial conditions for the (n+1)$^{th}$ application. This procedure was repeated until a steady state of the first oscillation was achieved. Usually less than 20 pump pulse applications were sufficient. This procedure was repeated for a large number of values for the repetition rate of the pump pulse, and the response curve of the magnetometer is depicted by the solid line in Fig. 5 and Fig. 6. There is a very good agreement between the experiment and theory of both the Bloch equation and a good qualitative agreement with the quantum model presented in Reference [9] and in the theory chapter of this paper. As expected, the width of the resonances vs repetition rate decreases as the repetition rate decreases.

**<u>Bloch equations formalism for the resonance curves</u>**

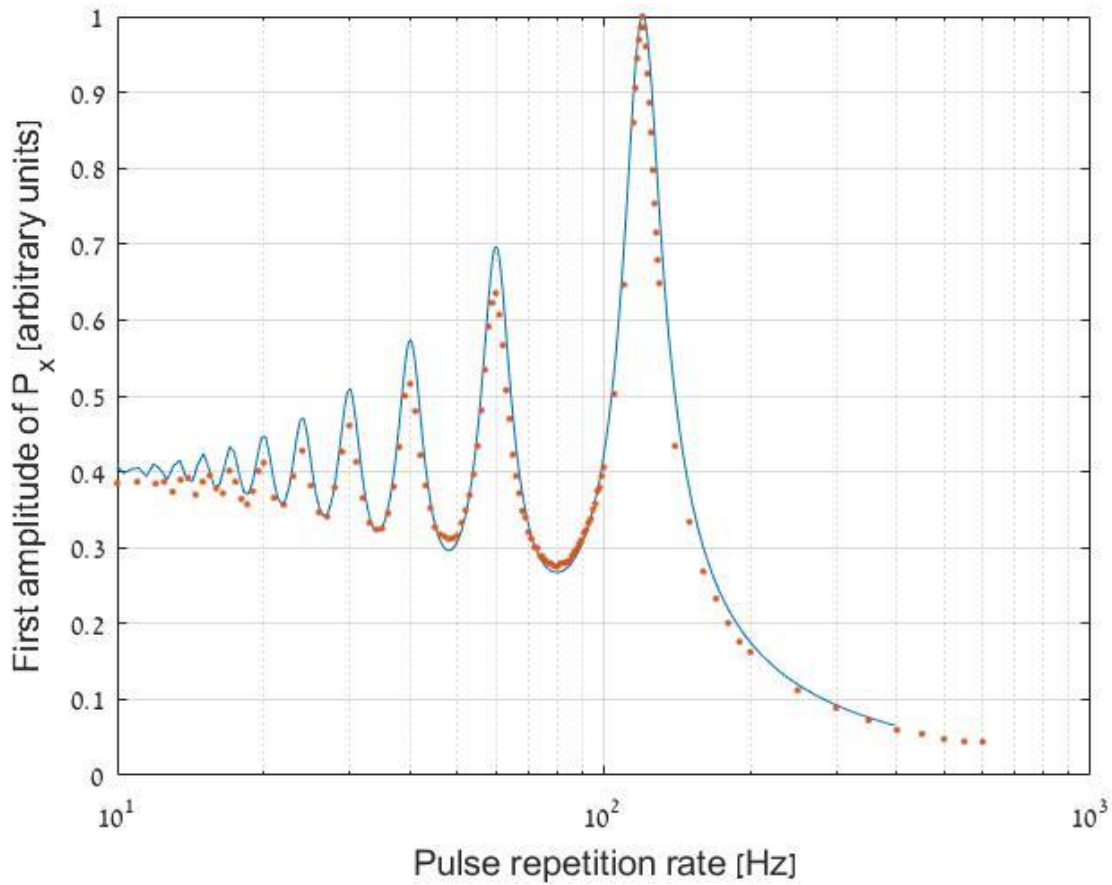

Fig. 5. Normalized resonance curves for 120Hz Larmor frequency for the K pumped Rb probed configuration. The dots represent the experimental points, the solid line shows the calculation results. The rate of spin destruction is calculated to be $\Gamma_{SD}=6.6$ sec$^{-1}$.

The Bloch equations describing the atomic system mentioned above are given in [8]. In the simple geometry of Fig. 1. The Bloch equation can be solved analytically. Using the pulsed pump CW probe scheme, one can reconstruct the waveforms depicted in Fig. 2. and calculate the amplitude of the first oscillation vs. the pulse rate of the pump laser. The procedure for which was described formerly. The analysis results are presented for two hybrid magnetometers: Potassium pumped and Rubidium probed and vice-versa. The results are given in Fig. 5 and Fig. 6.

The experiment was performed in the SERF regime. As mentioned by Savukov [3] in this regime the spin oscillation frequency is altered because of the "nuclear-slowdown-factor" and the oscillation frequency of the two isotopes of Rb, become the same, thus for the Rb probed experiments only one set of resonance frequencies is detected.

A distinguished feature of the measured resonance curve is the much larger width of the Rb probed resonance curves.

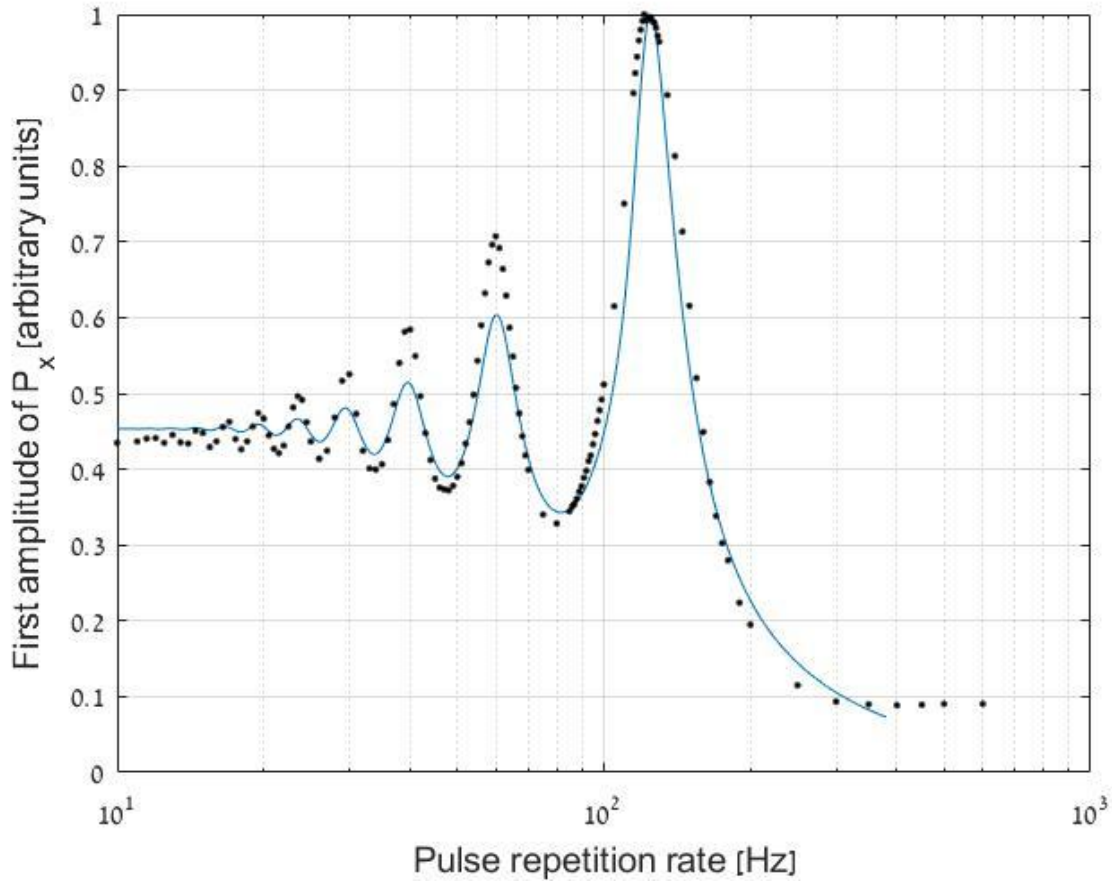

Fig. 6. Normalized resonance curves for 120Hz Larmor frequency for the Rb pumped K probed configuration. The dots represent the experimental points; the solid line shows the calculation results. The rate of spin destruction is calculated to be $\Gamma_{SD}=13$ sec$^{-1}$.

**Spin destruction rates of Potassium and Rubidium**

It is obvious from the experimental data and the analysis presented in Fig. 5 and Fig. 6 that the spin destruction (SD) rate for Rb pumped spin is much larger than that of K pumped spins. The alkali metals cell used in the experiment consisted of 2.5 amagats of $^4$He, 52 torr of N$_2$, and a mixture of K and Rb delivering a ratio of 1:20 of vapor density of Rb and K respectively at 180°C. Using the data given in Ref. [8] for the spin-destruction (SD) cross sections of K and Rb by the various atoms in the cell, one can conclude that the most significant contribution to the total spin destruction is collisions with He atoms. The cross section for this process is small ($\sim 10^{-25}$ cm$^2$) but the density of He atoms is of the order of $10^{20}$ cm$^{-3}$. The SD cross section for Rb by He is 9 times higher than that of K. The reduction of the spin destruction in the SERF regime by the nuclear-slowing-down factor, q, reduces this ratio by a factor of 2, (q for potassium is ~6, and for natural Rb[1] is ~10.2. Thus, the SD rate of Rubidium in our experiment should be higher by a factor of 4 than that of Potassium. Our measurements yield a factor of 2, in reasonable agreement with the phenomenological model prediction.

**Extended Theory**

A quantum theoretical three levels Λ Lambda scheme was studied, which presents a simplification of the active alkali atoms levels [10]. It shows that the presence of a sequential pulsed pump at a repetition rate equal to 1/n of any internal atomic frequency exhibits resonances. Expanded details of this model are given below. In optical magnetometer, magnetic interaction (μ·H) generates Zeeman splitting at the Larmor frequency and characteristic internal Zeeman frequency. Thus, one expects to see the above resonances when the pump is a sequential at repetition rate of f = 1/n the Larmor frequency. The present study exhibits this effect in the particular case of the Atomic magnetometer. We utilize the scalar mode of the optical magnetometer in a pulse mode excitation and measure the temporal response of the decaying magnetic moment or the population in the related Zeeman level [11]. Intuitively, this effect is a result of the synchronization of the pump rate with the internal eigenfrequency of the quantum system, the alkali atom in the present case of the Optical magnetometer. Although the repetition rate has no quantal characteristics, Quantum resonances occur due to the synchronization. Quantum calculations of three levels system model indicate possible false absorption measurement at resonance frequency employed in various magnetometry sensitivity level on the order of $10^{-8}$, due to possible EIT effect utilizing pump-probe in CW or pulsed modes. Methods in which the magnetic detection employs only frequency changes do not suffer from this effect [6].

In the sake of completeness, we briefly sketch here our previous work [9] in which an atomic SERF magnetometer was modelled as a three-level system driven by two lasers. The following system Hamiltonian has been used

$$H = \hat{d}_{ac} \cdot \vec{E}_{ac} f_1(t) + \hat{d}_{bc} \cdot \vec{E}_{bc} f_2(t) + \hat{d}_{ac} \cdot \vec{E}_{bc} f_2(t) e^{-i\omega_{ab}t} + \hat{d}_{bc} \cdot \vec{E}_{ac} f_1(t) e^{i\omega_{ab}t} + H.c., \quad (5)$$

where $\hat{d}_{ac} = d_{ac}\sigma_{ac}$ and $\hat{d}_{bc} = d_{bc}\sigma_{bc}$ are dipole operators of the corresponding transitions, and $\sigma_{ac} = |c\rangle\langle a|$ and $\sigma_{bc} = |b\rangle\langle a|$ are atomic raising operators. The first two terms in the Hamiltonian (7) represent resonant interaction between the driving fields $E_{ac}$ and $E_{bc}$ and the corresponding transitions. The last two terms describe cross interaction between the driving field $E_{ac}$ ($E_{bc}$) and the transition $|b\rangle \rightarrow |c\rangle$ ($|a\rangle \rightarrow |c\rangle$), in which case the fields are automatically detuned from the corresponding transitions by the atomic frequency $\pm\omega_{ab}$. Functions $f_1(t)$ and $f_2(t)$, of the type

$$f(t) = \sum_n \{g(t - n\tau)\} \quad (6)$$

which determine the time dependencies of the driving field amplitudes $E_{ac}$ and $E_{bc}$, respectively, are chosen either as *1*, in the case of cw field, or as $g(t - n\tau)$ in the case of a sequence of short pulses, or any combination thereof. The function *g* here is taken to be Gaussian, but any appropriate pulse function can be used as well. Solving numerically the master equation with the Hamiltonian (5) the coherences and level populations were calculated for various values of the system parameters. The dependence of the oscillation amplitude of Im[$\rho_{bc}$], which is responsible for absorption on the probe transition $|b> \rightarrow |c>$, on the repetition rate $\omega_{Rep}$, exhibits, apart of intuitively expected main resonance at $\omega_{Rep} = \omega_{ab}$, other resonances at fractional frequencies $\omega_{Rep} = \omega_{ab}/n$ with *n* being an integer number, which agrees well with experimental result shown in Fig. 7.

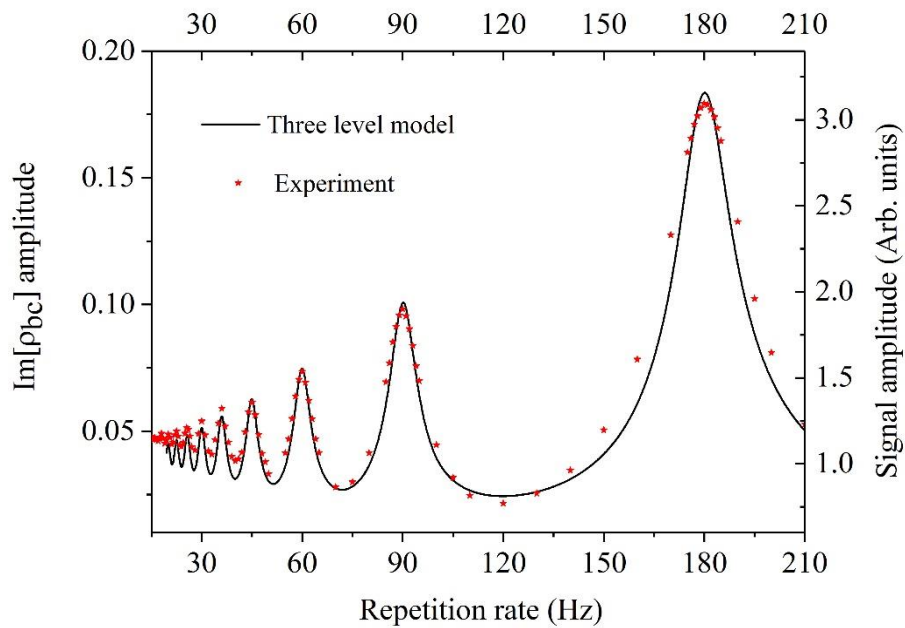

Fig. 7. Oscillation amplitude of the absorption on the probe transition as a function of the pulse repetition rate $\omega_{Rep}$ (solid line). Apart from expected resonance at $\omega_{Rep} = \omega_{ab}$ there are additional resonances at $\omega_{Rep} = \omega_{ab}/n$, where *n* is an integer number.

Further calculations show electromagnetically induced transparency (EIT) [10] and Coherent Population Trapping (CPT) shown in Fig. 8, as expected in a three levels system.

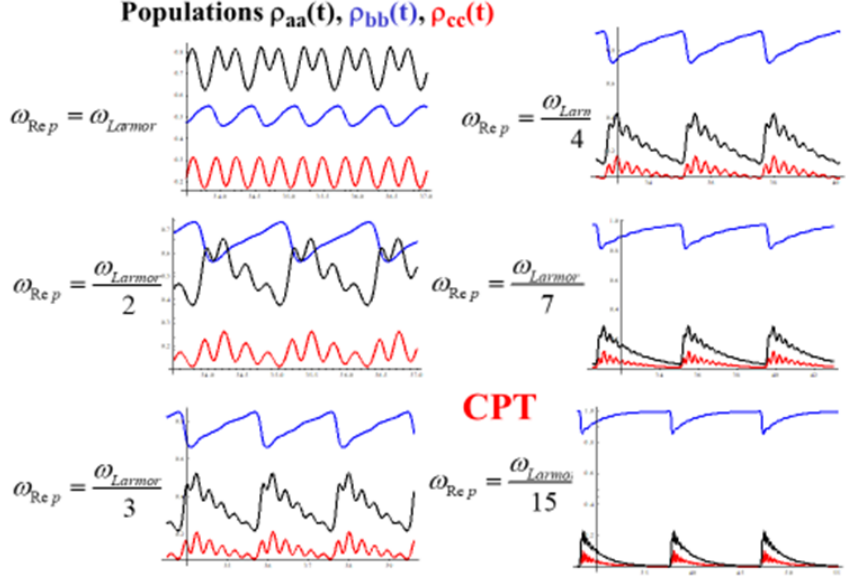

Fig. 8. Populations of atomic levels as functions of time at quantum 1/n resonances. There is clear CPT at fractional resonances $\omega_{Rep} = \omega_{ab}/n$, at $n > 3$.

It is important note for atomic magnetometer measurements that rely on measuring absorption intensity of the probe, particularly in the case where single laser is used for both pump and probe CW lasers. There is coherence among the pump and probe that may result experimentally in EIT. This is detrimental for the measure of the absorption amplitude, and in turn the detection sensitivity of a magnetometer. We note that EIT is readily observed in Alkali atoms [10]. This is obtained in our simplified theoretical calculation, Fig. 9. This may jeopardize the accuracy in such measurements in atomic magnetometers, as the probe intensity measurement may not reflect truly the change due to the magnetic field. This problem is not present in the case where magnetic field is measured solely from changes in the Larmor frequency rather than absorption intensity [6].

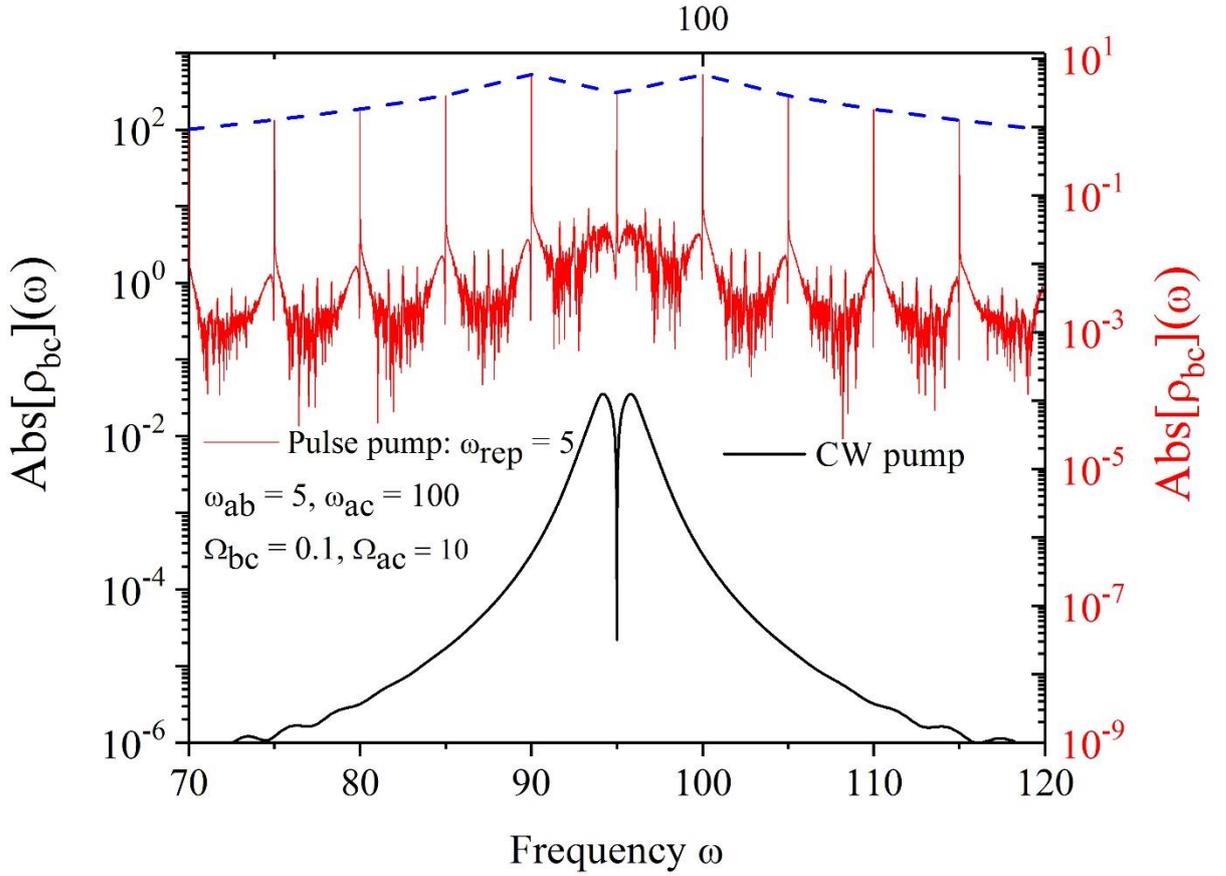

Fig. 9. Comparison of the probe absorption spectra for CW (bottom line) vs pulse (upper line) pump. EIT is clearly manifested in both cases. This may cause false positive sensitivity in optical magnetometer measurements of minute magnetic field, since absorption amplitude may adversely show lesser magnitude than it is in reality.

Another interesting result is obtained at a specific resonant condition when the Rabi frequency $\Omega_{ac} = \vec{d}_{ac} \cdot \vec{E}_{ac}/\hbar$ of the pump laser equals the Larmor frequency $\omega_{ab}$. In this case in addition to *1/n* resonances there appear more resonances at $\omega_{Rep} = 2\omega_{ab}/(2n+1)$, as can be seen in Fig. 10. These resonances are quite prominent due to resonance condition of $\Omega_{ac} = \omega_{ab}$.

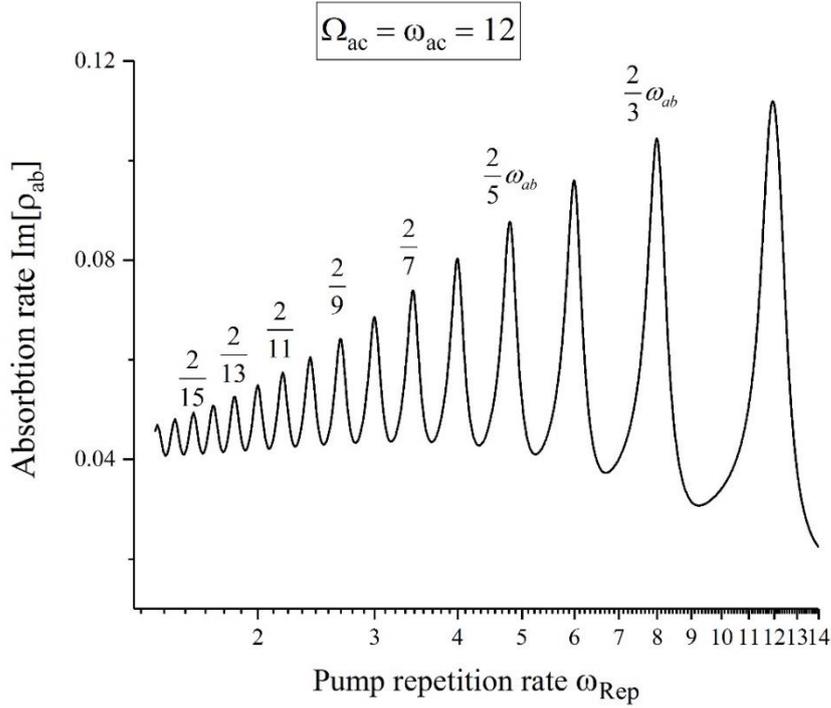

Fig. 10. Fractional resonances at $\omega_{Rep} = 2\omega_{ab}/(2n+1)$ under resonant condition $\Omega_{ac} = \omega_{ab}$.

Intuitively $2/(2n+1)$ peaks result from synchronization of populating $\rho_{cc}$. Population $\rho_{bb}$ of the intermediate level oscillates at Larmor frequency, which in this case is resonant to Rabi frequency $\Omega_{ac}$ of the pumping laser, the population $\rho_{cc}$ of the upper level oscillates at double frequency $2\Omega_{ac}$. This can be seen in Fig. 11.

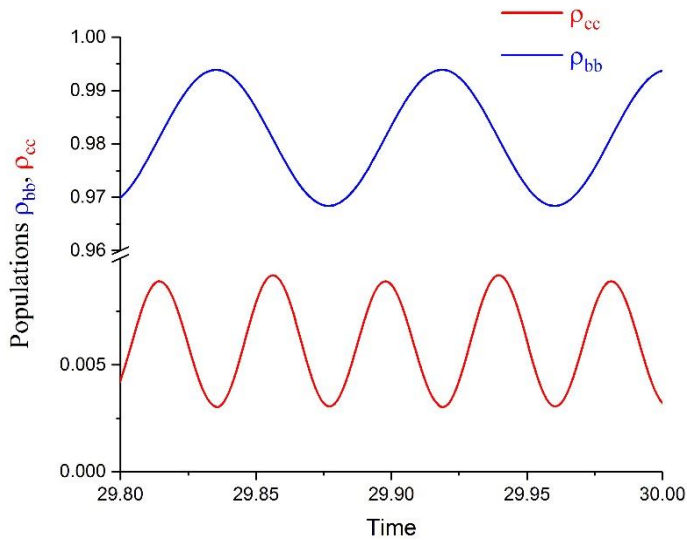

Fig. 11. While population $\rho_{bb}$ of the intermediate level oscillates at Larmor frequency, which in this case is resonant to Rabi frequency $\Omega_{ac}$ of the pumping laser, the population $\rho_{cc}$ of the upper level oscillates at double frequency $2\Omega_{ac}$.

In another condition the *2/(2n+1)* resonances are extremely weak and not observed experimentally.

## Summary

Operation of the optical magnetometer in a pulsed sequential mode at repetition rate at 1/n value of Larmor frequency of the active atom exhibits resonances as function of the pulse pump laser repetition rate. These resonances are in qualitative agreement with quantum theory of three levels Lambda system, pumped in a similar manner.

Interesting narrowing of these resonances is obtained when two alkali atoms are used in which one species acts as the pumping agent and transfers the spin excitation population to the other specie atom. By this method, isolation of the probe from the pump is achieved along with narrower resonances and higher "Q value". In particular, it is found that for this narrowing, the right choice is to use Potassium atom as the pump and the Rubidium atom as the probing one.

An important caution note is presented in this work for measurements of the magnetic field in methods that rely on absorption intensity to determine minute magnetic field changes. These are only reliable when there is no EIT (in CW pump) or absorption dip (in pulsed) effects. The method employing solely measuring changes in the Larmor frequency does not suffer from such problem.

## Acknowledgement

This research is supported in part by ONRG GRANT - NICOP - N62909-19-1-2030.